\documentclass[aps,prl,twocolumn,superscriptaddress]{revtex4-1}

\usepackage{graphicx}
\usepackage{bm}
\usepackage{color}
\usepackage{amsmath}
\usepackage{soul}

\begin{document}

\def\mos2{$\mathrm{Mo S_2}$}
\def\mote2{$\mathrm{Mo Te_2}$}
\def\wse2{$\mathrm{W Se_2}$}
\def\wte2{$\mathrm{W Te_2}$}

\title{Controlling the quantum spin Hall edge states in two-dimensional transition metal dichalcogenides}

\author{Artem Pulkin}
\email{Present address: QuTech and Kavli Institute of Nanoscience, Delft University of Technology, 2600 GA Delft, The Netherlands \\
E-mail: gpulkin@gmail.com}
\affiliation{Institute of Physics, Ecole Polytechnique F\'{e}d\'{e}rale de Lausanne (EPFL), CH-1015 Lausanne, Switzerland}

\author{Oleg V. Yazyev}
\email{E-mail: oleg.yazyev@epfl.ch}
\affiliation{Institute of Physics, Ecole Polytechnique F\'{e}d\'{e}rale de Lausanne (EPFL), CH-1015 Lausanne, Switzerland}
\affiliation{National Centre for Computational Design and Discovery of Novel Materials MARVEL, Ecole Polytechnique F\'{e}d\'{e}rale de Lausanne (EPFL), CH-1015 Lausanne, Switzerland}

\date{\today}

\def\artem#1{\textcolor{red}{#1}}

\begin{abstract}
    Two-dimensional  transition metal dichalcogenides (TMDs) of Mo and W in their 1T' crystalline phase host the quantum spin Hall (QSH) insulator phase.
    We address the electronic properties of the QSH edge states by means of first-principles calculations performed on realistic models of edge terminations of different stoichiometries.
    The QSH edge states show a tendency to have complex band dispersions and coexist with topologically trivial edge states.
    We nevertheless identify two stable edge terminations that allow isolating a pair of helical edge states within the band gap of TMDs, with monolayer 1T'-\wse2 being the most promising material.
    We also characterize the finite-size effects in the electronic structure of 1T'-\wse2 nanoribbons.
    Our results provide a guidance to the experimental studies and possible practical applications of QSH edge states in monolayer 1T'-TMDs.
\end{abstract}


\maketitle

\noindent

\begin{figure}
    \includegraphics{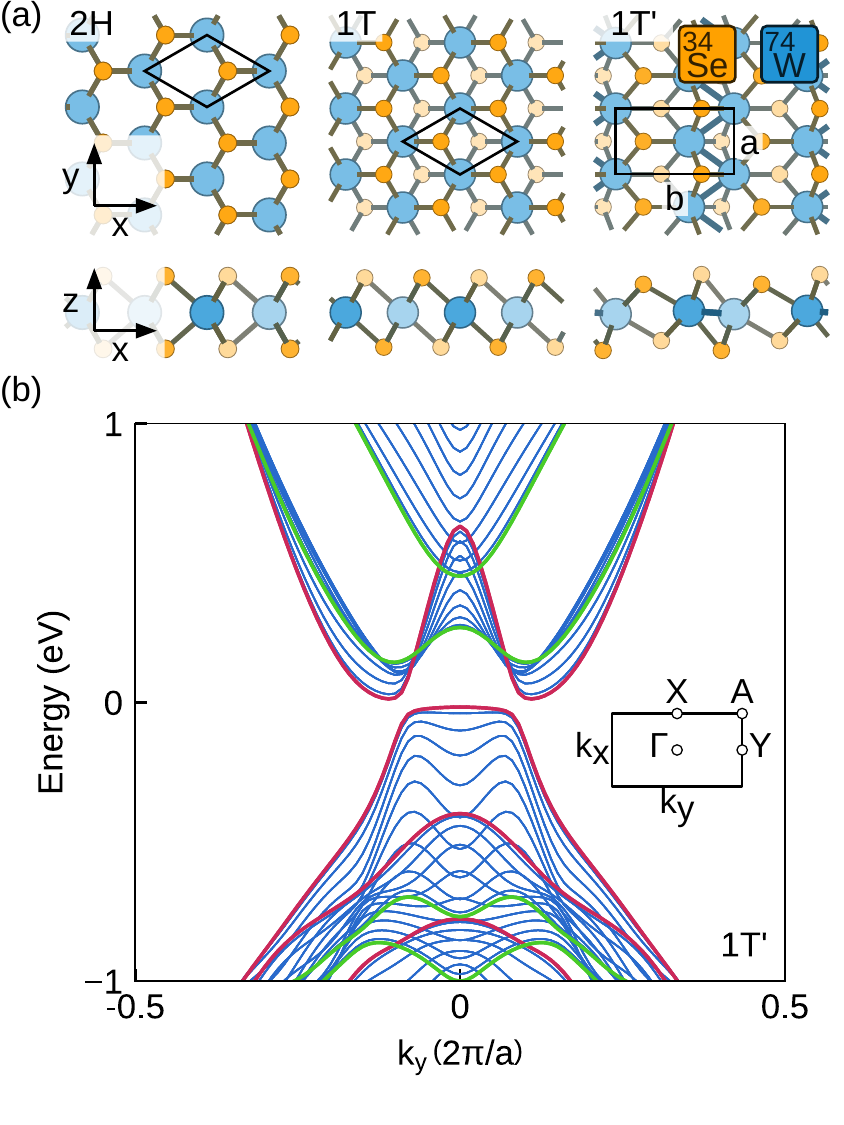}
    \caption{
(a) Atomic structure of the 2H, 1T and 1T' phases of monolayer \wse2 shown from two perspectives. The unit cells and lattice constants for the 1T'  phase are indicated.
(b) Band structure of monolayer 1T'-\wse2 plotted along a series of equidistant lines along the $k_y$ direction.
Red and green lines correspond to $k_x =0$ (Y--$\Gamma$--Y path) and  $k_x =\pi/b$ (A--X--A path). The Fermi level is inside the band gap.
    }
    \label{fig:1}
\end{figure}

Two-dimensional transition metal dichalcogenides (2D TMDs) of chemical composition $MX_2$, where $M$ is a transition metal element and $X$~=~S,~Se or Te, is a broad family of emerging 2D materials that attracts ever increasing interest~\cite{wang_electronics_2012, chhowalla_chemistry_2013, manzeli_2d_2017}.
Among them, group VI TMDs ($M$~=~Mo,~W) in their 2H crystalline phase continue receiving special attention due to their semiconducting band gaps that span a broad interval of energies~\cite{mak_atomically_2010, splendiani_emerging_2010}, which makes them suitable for diverse applications that include electronics and optoelectronics~\cite{radisavljevic_single-layer_2011}, photovoltaics~\cite{bernardi_extraordinary_2013,ugeda_giant_2014} and sensors devices~\cite{perkins_chemical_2013}.
Strong spin-orbit interactions in these materials~\cite{zhu_giant_2011} results in an intrinsic spin-valley coupling~\cite{xiao_coupled_2012} responsible for novel transport properties~\cite{habe_spin-dependent_2015, pulkin_spin-_2016}, the exotic Ising superconductor phase~\cite{lu_evidence_2015}, and more.
While the 2H phase is the thermodynamically stable  polymorph for all group VI TMDs except WTe$_2$, the metastable 1T phase has also been widely documented~\cite{voiry_phase_2015}.
The 1T phase is prone to lattice instabilities~\cite{duerloo_structural_2014, pasquier_notitle_nodate} giving rise to the so-called 1T' phase that features dimer-chain distortions observed in experiments \cite{chou_controlling_2015, gao_atomic-scale_2015, keum_bandgap_2015, cho_phase_2015, jia_direct_2017, peng_observation_2017}. The structures of these polymorphs of 2D TMDs are summarized in Figure~\ref{fig:1}a.

The 1T' crystalline polymorph of group VI monolayer TMD materials was theoretically predicted to realize the quantum spin Hall (QSH) insulator phase \cite{qian_quantum_2014}.
This topological electronic phase is due to the combination of band inversion and spin-orbit interactions that open few tens millielectronvolt band gaps \cite{kane_$z_2$_2005, choe_understanding_2016}.
A representative electronic band structure of monolayer 1T'-\wse2 is shown in Figure~\ref{fig:1}b.
The predicted QSH insulator phase soon received experimental confirmation in monolayer 1T'-\wte2 using both spectroscopic techniques and transport measurements \cite{tang_quantum_2017, fei_edge_2017}.
More recently, the same topological phase was observed in metastable monolayer 1T'-\wse2 further extending this family of 2D topological insulators \cite{chen_large_2018, ugeda_observation_2018}.

The most important consequence of the QSH insulator phase in material's bulk is the presence of helical edge states at its boundaries. 
The properties of the topological edge states in 2D TMDs are extensively discussed \cite{qian_quantum_2014, tang_quantum_2017, ugeda_observation_2018, wu_observation_2018, jelver_spontaneous_2019}
in the context of unconventional electronics and quantum computing due to their unique properties: formation of exotic quasiparticles, spin-momentum locking, protection against backscattering, to name just a few \cite{kitaev_periodic_2009, hasan_colloquium_2010, qi_topological_2011}.
While the existence of these topological edge states is guaranteed by bulk topology, the details of their band dispersion and spin texture would depend on the local structure of the edges, such as the crystallographic orientation, chemical composition and termination\cite{wang_tuning_2014}.
More importantly, the simultaneous presence of undesired topologically trivial edge states is not guaranteed by any existing argument.
Thus, establishing the relations between the structure of the edges and the properties of QSH edge states is of crucial importance for advancing the fundamental physics and realizing technological applications of these topological phases.

In this Letter, we address the properties of the quantum spin Hall edge states in monolayer 1T'-phase transition metal dichalcogenides from first principles.
We consider realistic edge terminations of different chemical composition identifying stable configurations.
Our non-equilibrium Green's function (NEGF) calculations reveal predominantly complex dispersions of the QSH edge states coexisting with topologically trivial edge states. Two edge terminations, however, were shown to host an isolated pair of helical states. We also address finite-size effects in the electronic structure of 1T'-\wse2 nanoribbon models providing further guidance to the experimental studies of these novel topological states.

Our first-principles electronic structure calculations have been performed using the OpenMX software package within the framework of density functional theory (DFT) and  localized pseudoatomic orbital basis sets \cite{ozaki_variationally_2003, ozaki_numerical_2004}.
The basis set used in our work contains 25 basis functions per transition metal atom (s2p2d2f1) and 14 basis functions per chalcogen atom (s3p3d1), which is sufficient for accurate description of the electronic structure. The calculations have been performed using the PBE exchange-correlation functional \cite{perdew_generalized_1996} and taking into account the spin-orbit interaction.
The relaxed atomic structures of the edges were obtained in the nanoribbon geometry of width up to 5.5~nm with two equivalent edges.
The electronic structure of the edges was then addressed using semi-infinite models combined with the tight-binding Kohn-Sham Hamiltonian and overlap matrices obtained from DFT calculations.
The local density of electronic states was calculated using the surface Green's function approach.

\begin{figure}
    \includegraphics{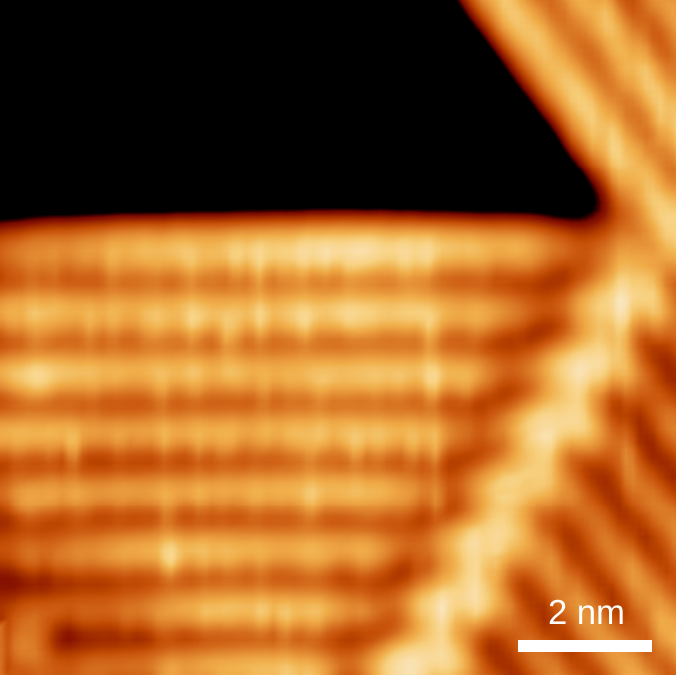}
    \caption{
        Scanning tunnelling microscopy (STM) image of monolayer 1T'-\wse2 showing ordered edges oriented along the dimer rows.
        STM imaging parameters and other experimental details can be found in Ref.~\cite{pedramrazi_manipulating_2019}.
    }
    \label{fig:stm}
\end{figure}

\begin{figure*}
    \includegraphics{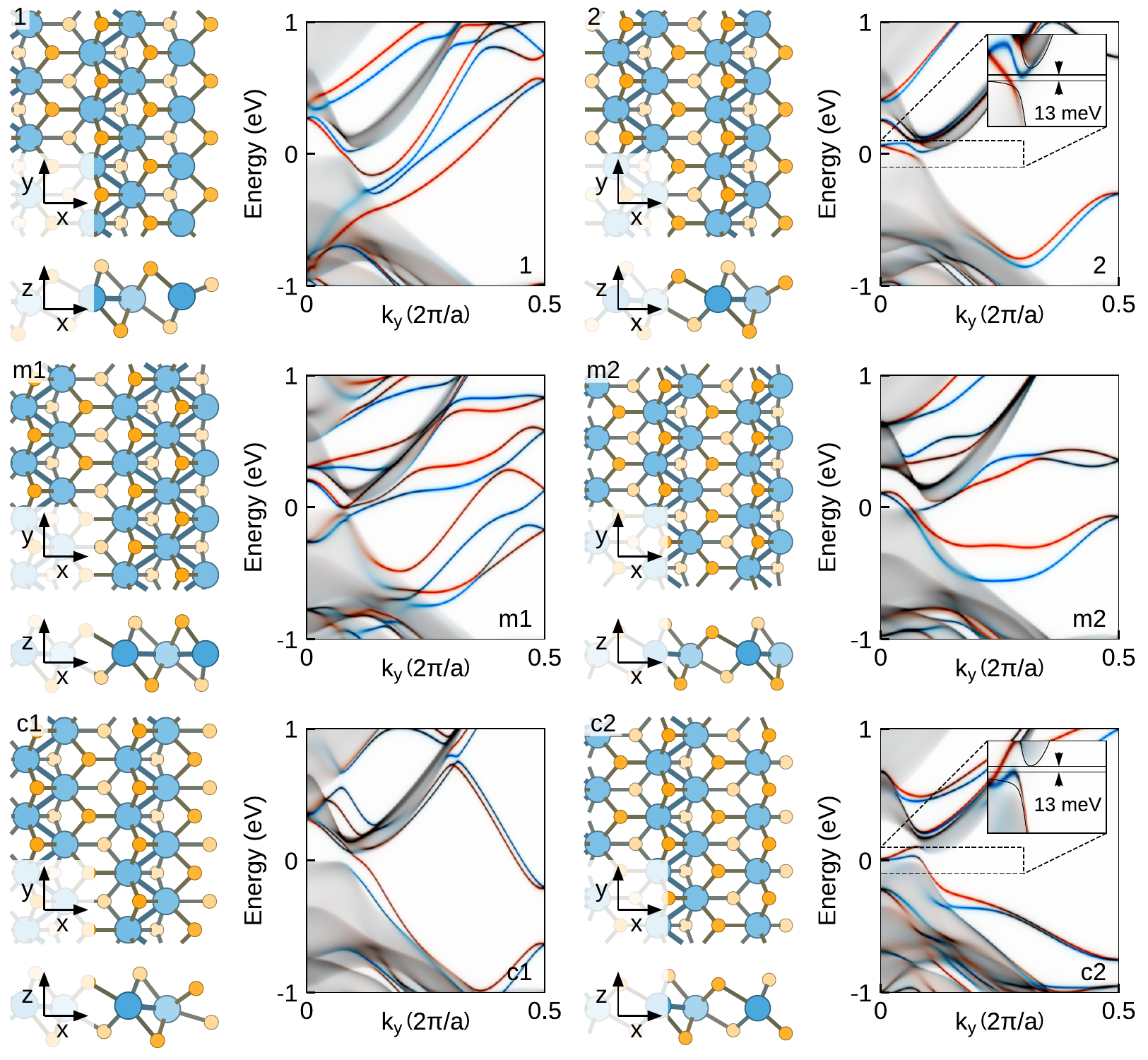}
\caption{Possible first-principles models of six terminations of monolayer 1T'-\wse2: stoichiometric (1, 2), metal-rich (m1, m2) and chalcogen-rich (c1, c2) together with the corresponding calculated momentum- and spin-resolved local densities of states.
The atomic structures of the edges are shown from two perspectives.
$E = 0$ corresponds to the Fermi energy and half of the one-dimensional Brillouin zone is displayed.
The gray areas correspond to the projection of spin-degenerate bulk states, while the edge states are depicted in red and blue for spin-up and spin-down channels, respectively.
The insets for terminations 2 and c2 show the details of edge-state dispersion within the band gap of monolayer 1T'-\wse2.
In these insets, the contours of the projected bulk bands are shown as solid curves.
The horizontal lines delimit the energy ranges in which a single pair of QSH edge states is isolated.
}
    \label{fig:2}
\end{figure*}

\begin{figure}
    \includegraphics{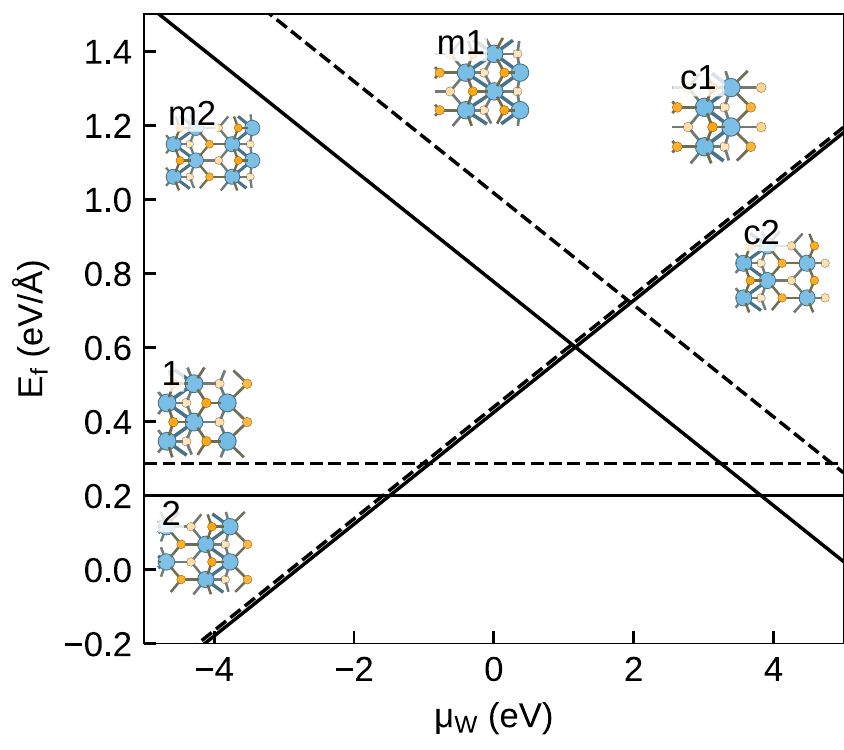}
\caption{Formation energies $E_f$ of the six considered edges terminations of monolayer 1T'-\wse2 as a function of the chemical potential of tungsten $\mu_\mathrm{W}$. The reference chemical potential
$\mu_\mathrm{W} = 0$ corresponds to bulk bcc tungsten.}
    \label{fig:nrg}
\end{figure}

Even though our work addresses all six TMDs involving group VI transition metals, we will focus our discussion on \wse2 as a representative example for the reasons explained below.
Fig.~\ref{fig:1}d shows the electronic band structure of monolayer 1T'-\wse2. The calculated band gap of 29~meV agrees well with previous studies performed at the DFT level~\cite{qian_quantum_2014, pulkin_robustness_2017}, while 
the evident flattening of the valence band is indicative of band inversion responsible for the QSH phase.
The presence of the topological phase gives rise to the helical states localized at the interface with a topologically trivial medium \cite{kane_$z_2$_2005}, with the simplest example of such an interface being the edge.
The edges can be characterized by various crystallographic orientations and terminations, with the ones corresponding to the lowest formation energies being thermodynamically preferred.
The crystallographic orientation of the edges is defined by the periodicity vector ${\mathbf{d}} = n  {\mathbf{a}} + m  {\mathbf{b}} \equiv (n,m)$, where $\mathbf{a}$, $\mathbf{b}$ are the lattice vectors of the 1T'-TMD monolayer, see Fig.~\ref{fig:1}(a).
Below, we will limit our consideration to the shortest possible ${\mathbf{d}} = (1,0)$  that corresponds to the ``zigzag'' direction of the underlying honeycomb lattice.
Such edges are prevalent in the CVD-grown 2H-phase TMDs~\cite{van_der_zande_grains_2013,zhou_intrinsic_2013,dumcenco_large-area_2015} as well as in monolayer 1T'-\wte2~\cite{peng_observation_2017}.
More recently, the tendency to form well-ordered edges was also observed in 1T'-\wse2 by means scanning tunneling miscroscopy\cite{pedramrazi_manipulating_2019}.
Figure \ref{fig:stm} reproduces one of experimental images of such edges coexisting with a 1T'-1T' domain boundary.
The dimer rows appear to be aligned parallel to the edges, and we will reproduce this feature in our models.
We consider six different terminations of the edges shown in Fig.~\ref{fig:2}, which can be divided into three groups (rows in Fig.~\ref{fig:2}) according to the local chemical composition -- stoichiometric (1 and 2), metal-rich (m1 and m2) and chalcogen-rich (c1 and c2). The two configurations in each pair differ by the position of the dimer chains relative to the edge.
Since TMDs have binary chemical composition, the formation energies $E_f$ of the edges depend on the chemical potentials of the elements.
The former is defined per unit length as
\begin{equation}
E_f = \left ( E_\mathrm{ribbon} - N_\mathrm{W} \mu_\mathrm{W} - N_\mathrm{Se} \mu_\mathrm{Se} \right ) / 2 a ~,
\end{equation}
where $E_\mathrm{ribbon}$ is the total energy of a \wse2 nanoribbon model with two identical edges, $N_\mathrm{W}$ and $N_\mathrm{Se}$ are the numbers of W and Se atoms in the model, respectively, $\mu_\mathrm{W}$ and $\mu_\mathrm{Se}$ are the corresponding chemical potentials and $a$ is the lattice constant in Fig.~\ref{fig:1}.
The chemical potentials of the two elements are subject to the bulk-phase energy $E_\mathrm{bulk}$ constraint
\begin{equation}
E_\mathrm{bulk} = 2 \mu_\mathrm{W} + 4 \mu_\mathrm{Se} = \mathrm{const} ~.
\end{equation}
The formation energies of terminations 1, m1 and c1 in Fig.~\ref{fig:2} differ from those of terminations 2, m2 and c2, respectively, by a constant since in each pair of models the edges have the same chemical composition.
Figure \ref{fig:nrg} shows the formation energies $E_f$ all six edge configurations for monolayer 1T'-\wse2 as a function of $\mu_\mathrm{W}$ with $\mu_\mathrm{W} = 0$ that corresponds to bulk bcc tungsten.
Three different terminations (2, m2 and c2) can be identified as the most stable within distinct ranges of the chemical potential.
Other monolayer 1T'-TMDs show a similar behavior, in particular, the metal-rich structure m1  is always energetically unfavorable suggesting that the unbound chain of metal atoms at the edge has a considerable energy overhead. The corresponding plots of  formation energies as a function of the chemical potential of metal are presented in the Supplementary Material document. We would also like to point out that all investigated edge terminations for all six TMDs do not show any local magnetic moments at the edge, which is of crucial importance for exploiting the topological protection of the QSH edge states.  
This result contrasts to a recent investigation that finds magnetic edge configurations for monolayer 1T'-\mos2 edges \cite{jelver_spontaneous_2019}.

The electronic structure of the considered six configurations is addressed by plotting the momentum- and spin-resolved local density of states at the edge, shown in half of the Brillouin zone in Fig.~\ref{fig:2}.
Bulk electronic states of monolayer 1T'-TMDs are spin-degenerate due to the presence of time-reversal and inversion symmetries. However, inversion symmetry is broken at the edges, hence spin-degeneracy of the edge states is expected to be lifted. This allows to unambiguously distinguish bulk and edge states in the discussed local density of states plots. Bulk states appear as gray areas that correspond to the projections of the 2D bulk band structure onto the momentum $k_y$ parallel  to the edge. The shape of these areas does not depend on the local atomic structure of the edges. In contrast, edge states are spin-polarized (in Fig.~\ref{fig:2} color-coding reflects the expectation value of the $s_z$ operator, with red and blue corresponding to spin-up and spin-down channels, respectively) and appear as band features that vary strongly upon changing the edge termination.
The spin-polarized bands cross the Fermi level an odd number of times as expected for the topologically non-trivial QSH phase in monolayer 1T'-\wse2. In general, the dispersion of edge states in monolayer 1T'-\wse2 and other 1T'-TMDs is very complex as compared to the idealized picture of two linearly dispersing bands forming a single crossing.
The number of localized modes inside the band gap varies strongly and may be as large as 7 per spin (terminations 1 and m1). These bands can be viewed as the topological QSH edge states combined with the topologically trivial edge states that show a typical Rashba-like dispersion~\cite{bychkov_notitle_1984}. 
Interestingly, the out-of-plane spin-polarization of the edge bands is not always preserved across the entire Brillouin zone.
For example, in the case configuration m2 two edge states  form an avoided crossing due to the spin-orbit coupling at which spin mixing takes place and hence the spinor wavefunctions are no longer eigenstates of the $s_z$ spin operator.

The topological protection of ballistic transport of charge carriers refers to a process in which charge carriers are supported by one of the two helical edge modes of the QSH insulator without backscattering. This fundamental property of the topological edge states is interesting from the point of view of exploring novel physical phenomena as well as technological applications. One often quoted idea is the use of the QSH edge channels as  interconnect in nanoelectronic devices that operate in
dissipationless ballistic transport regime. These applications, however, depend critically on the ability to isolate the QSH edge states from bulk states as well as topologically trivial edge states within the gap that may enable scattering processes.
We therefore focus on identifying the combination of edge termination and material composition that results in the presence of a single pair of QSH edge states within the gap and not accompanied by additional topologically trivial bands. According to our calculations, this condition is satisfied only for edge terminations 2 and c2 in monolayer 1T'-\wse2. The dispersion of edge-state bands in other studied TMDs is qualitatively similar (see Supplementary Material), but 1T'-\wse2 shows the broadest energy ranges in which a single pair of QSH helical edge states is isolated. Insets in Fig.~\ref{fig:2} show that in both cases the QSH edge states are isolated within the energy range of 13~meV inside the 29~meV band gap of monolayer 1T'-\wse2. In the case of stoichiometric termination 2 this energy window is pinned to the valence band maximum, while in the case of chalcogen-rich termination c2 it is located at the conduction band minimum of monolayer 1T'-\wse2. We thus conclude that this particular material and these two edge terminations present the most favorable combination for addressing the QSH physics and potential technological applications among the 2D TMD topological insulators.
For the same reason, our discussion focuses on this representative of the 1T'-phase monolayer TMD family of materials. 

\begin{figure}
    \includegraphics{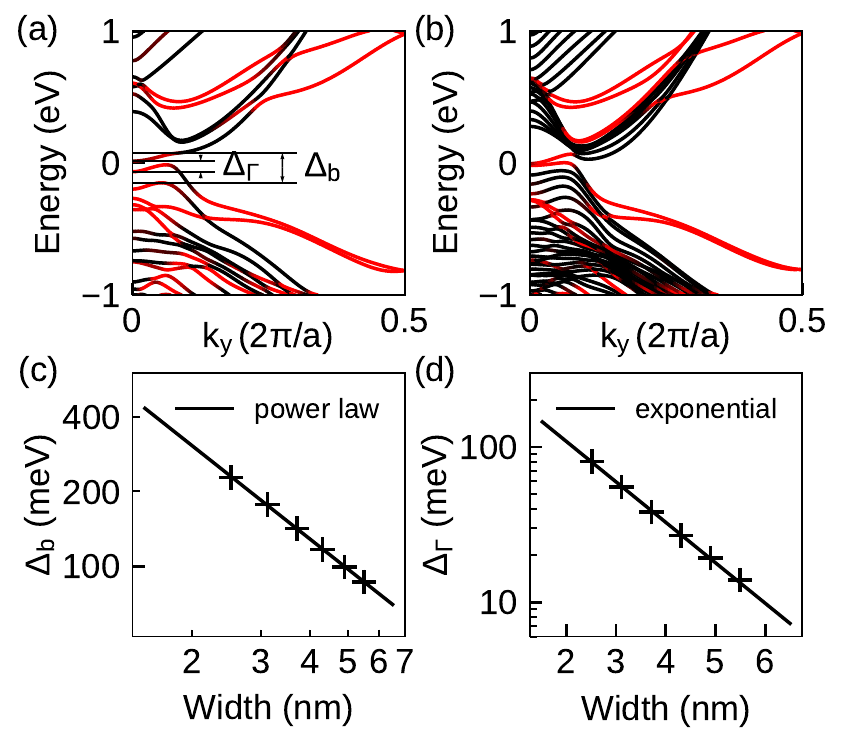}
\caption{Electronic band structures of monolayer 1T'-\wse2 nanoribbons with c2 edge termination of (a) 2.5~nm and (b) 5.5~nm width.
The bulk-like states are shown in black while the localized edge states are presented in red.
(c) Dependence of the bulk band gap $\Delta_b$ on the nanoribbon width $w$ for the above termination. The line shows a power law fit.
(d) Dependence of the $\Gamma$-point edge-state splitting on the nanoribbon width $w$. The line shows an exponential fit.
}
    \label{fig:3}
\end{figure}

We will now address the finite-size effects on the bulk and edge states of monolayer 1T'-\wse2 by considering nanoribbons of different width.
The interplay of the following two effects is expected to govern the electronic structure of nanoribbons upon decreasing width: (i) increase of the bulk band gap $\Delta_b$ due to the quantum confinement effects, and (ii) 
hybridization of the edge states localized at the opposite edges due to a finite wavefunction overlap, which results in a gap opening $\Delta_\Gamma$ at the $\Gamma$ point where the edge-state bands cross \cite{huang_understanding_2013} (see Fig.~\ref{fig:3}a for definitions).
To understand how these two quantities depend on the nanoribbon width we performed DFT calculations  of monolayer 1T'-\wse2 nanoribbons up to width $w = 5.5$~nm assuming the c2 termination of both edges.
For each nanoribbon model, we calculated the band structure in the 1D Brillouin zone and analyzed the spatial distribution of the Bloch states.
This allowed us to assign the states either bulk or edge character, shown in black and red in Figs.~\ref{fig:3}a,b for two representative nanoribbons of 2.5 nm and 5.5 nm width, respectively.
The dispersion of edge states clearly resembles the results of calculations performed on semi-infinite models (Fig.~\ref{fig:2}) with the bulk-like states forming a series of sub-bands due to the quantum confinement.
The bulk band gap $\Delta_b$ was calculated by considering only those states having a significant orbital contribution originating from the nanoribbon bulk.
Figs.~\ref{fig:3}c,d summarize the results for both $\Delta_b$ and $\Delta_\Gamma$ as a function of nanoribbon width $w$.
The effect of quantum confinement on the bulk band gap can be understood as a result of the quantization of the Bloch wavevector perpendicular to the nanoribbon edge.
The band structure of 1D nanoribbons is composed of the states of the 2D band structure of bulk material with
\begin{equation}
k_x = \pm \frac{\pi n}{N b},
\end{equation}
where $N b$ is the nanoribbon width given in terms of the corresponding lattice constant $b$, and $n \leq N$ is a positive integer.
The band gap in monolayer 1T'-\wse2 occurs between the band extrema located at $k_x = 0$, which are never satisfied by the above condition.
Thus, the size of the bulk band gap of nanoribbons is expected to be always larger than the gap of the parent bulk material.
In the limit of infinite width $w$ the nanoribbon gap converges to the bulk band gap.
Correspondingly, the values of $\Delta_b$ shown in Fig.~\ref{fig:3}c can be described by a power law that depends the particularities of the dispersion of bulk bands close to the band gap edges.
Numerical fit of the results to $\Delta_b \left ( w \right ) = c w^{-p}  + \Delta_b \left ( \infty \right) $
with $\Delta_b \left ( \infty \right) = 29 \mathrm{meV}$ being fixed to the bulk band gap of monolayer 1T'-\wse2 yields $p=1.25$ (line in  Fig.~\ref{fig:3}c).
The obtained exponent  differs significantly from $p=2$ expected for the ideal quadratic bands. We ascribe this difference to the peculiar shape of bulk bands strongly affected by band inversion in this material.

Unlike $\Delta_b$ governed by the bulk properties, the gap opening $\Delta_\Gamma$ at the $\Gamma$ point is rather defined by the exponential decay of edge states.
The values calculated for monolayer 1T'-\wse2 nanoribbons shown in Fig.~\ref{fig:3}d can be accurately fitted  by
\begin{equation}
    \Delta_\Gamma \left ( w \right ) = c e^{-w / \lambda} ~
\end{equation}
with $ c = 0.73$~eV and a characteristic decay length of edge states $\lambda = 1.7$~nm.
This relatively fast exponential decay combined with a slower polynomial decay of the bulk band gap suggests  a regime, in which quantum confinement can be used to enhance the bulk band gap of the topological insulator phase with no significant hybridization of the edge states.

To summarize, our work reveals that the dispersion of topological QSH edge states and the presence of trivial edge states within the band gap of monolayer 1T'-phase transition metal dichalcogenides depends strongly on the stoichiometry and termination of the edges. We identify edge terminations that are stable and allow isolating a pair of helical edge states within the band gap. Focusing on monolayer 1T'-\wse2 as a representative example we address also the finite-size effects in the electronic structure, thus providing a guidance to the experimental studies and possible practical applications of QSH edge states in monolayer 1T'-TMDs.

{\bf Acknowledgments.}
We acknowledge support by the ERC Starting grant “TopoMat” (Grant No. 306504) and NCCR Marvel.
Author A.P. thanks Swiss NSF for the support provided through Early Postdoc.Mobility program (project P2ELP2\_175281).
First-principles calculations were performed at the Swiss National Supercomputing Centre (CSCS) under project s832 and the facilities of Scientific IT and Application Support Center of EPFL. 
We thank C. Herbig, Z. Pedramrazi and M. F. Crommie for providing the STM image reproduced in Figure \ref{fig:stm}.

\bibliography{ms}

\end{document}


\title{Supplementary Material for: \\
``Controlling the quantum spin Hall edge states in two-dimensional transition metal dichalcogenides''}

\author{Artem Pulkin}
\affiliation{Institute of Physics, Ecole Polytechnique F\'{e}d\'{e}rale de Lausanne (EPFL), CH-1015 Lausanne, Switzerland}

\author{Oleg V. Yazyev}
\affiliation{Institute of Physics, Ecole Polytechnique F\'{e}d\'{e}rale de Lausanne (EPFL), CH-1015 Lausanne, Switzerland}
\affiliation{National Centre for Computational Design and Discovery of Novel Materials MARVEL, Ecole Polytechnique F\'{e}d\'{e}rale de Lausanne (EPFL), CH-1015 Lausanne, Switzerland}

\maketitle

\begin{figure*}
\includegraphics[width=17cm]{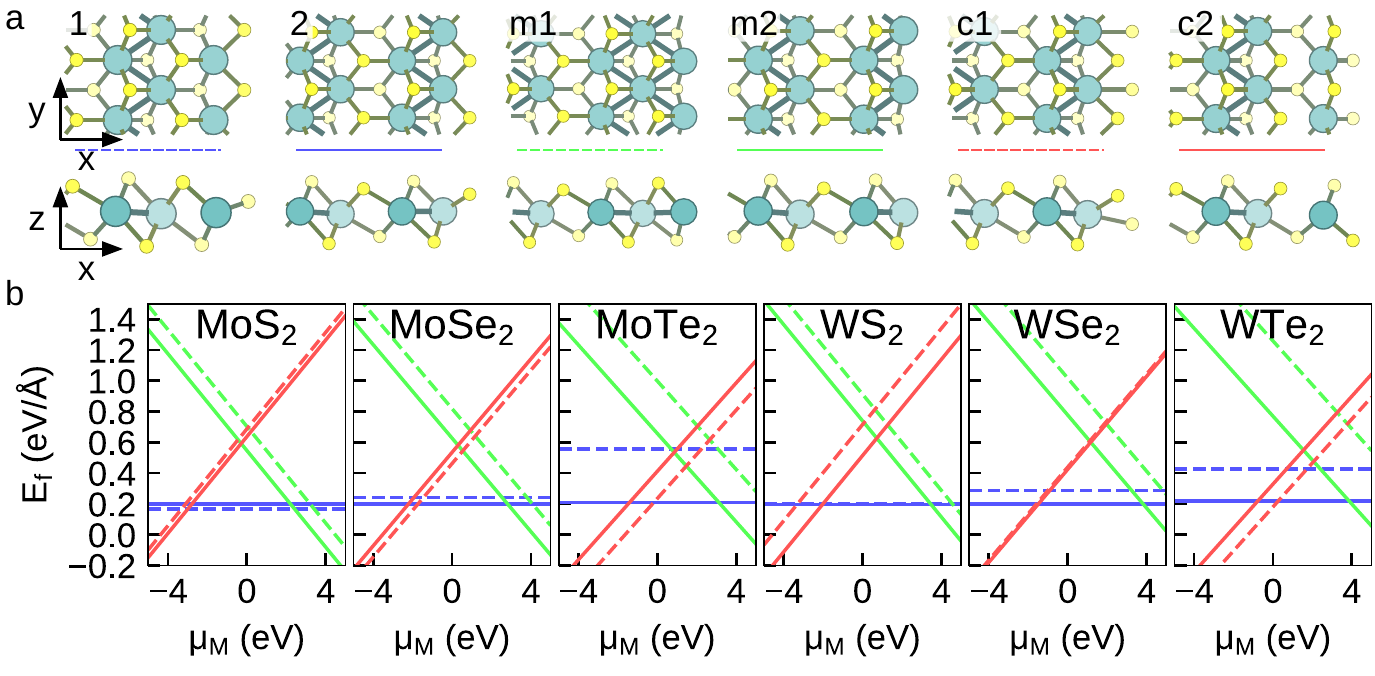}
\caption{
Formation energies $E_f$ of the six considered edges terminations of monolayer 1T'-TMDs as a function of the chemical potential of transition metal $\mu_\mathrm{M}$. Panel (a) shows the atomic structure of the edges. Panel (b) shows the formation energy plots.
The reference $\mu_\mathrm{M} = 0$ corresponds to the bulk bcc molybdenum and tungsten, respectively.
The formation energy $E_f$ is given per lattice constant of the corresponding material.
}
\label{fig:s1}
\end{figure*}

\newpage

%

\begin{figure*}
\includegraphics[width=17cm]{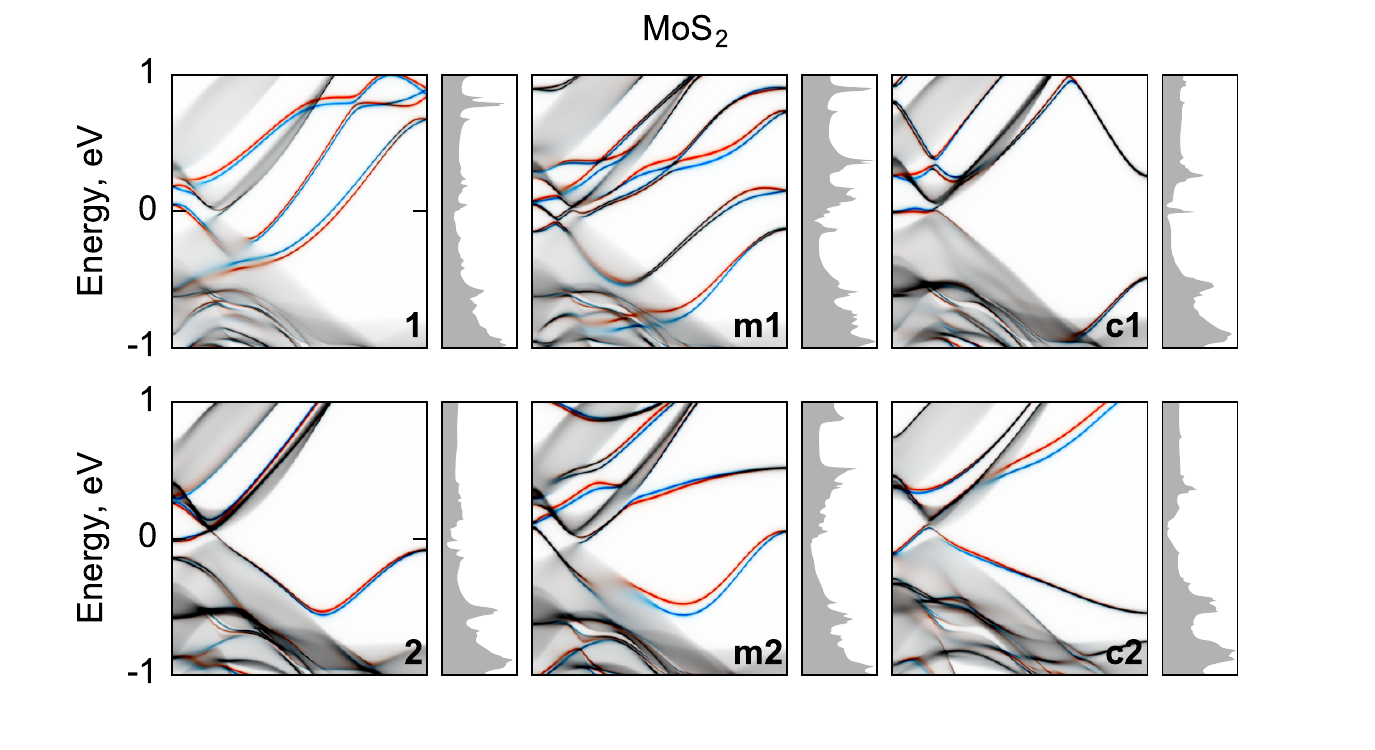}
\includegraphics[width=17cm]{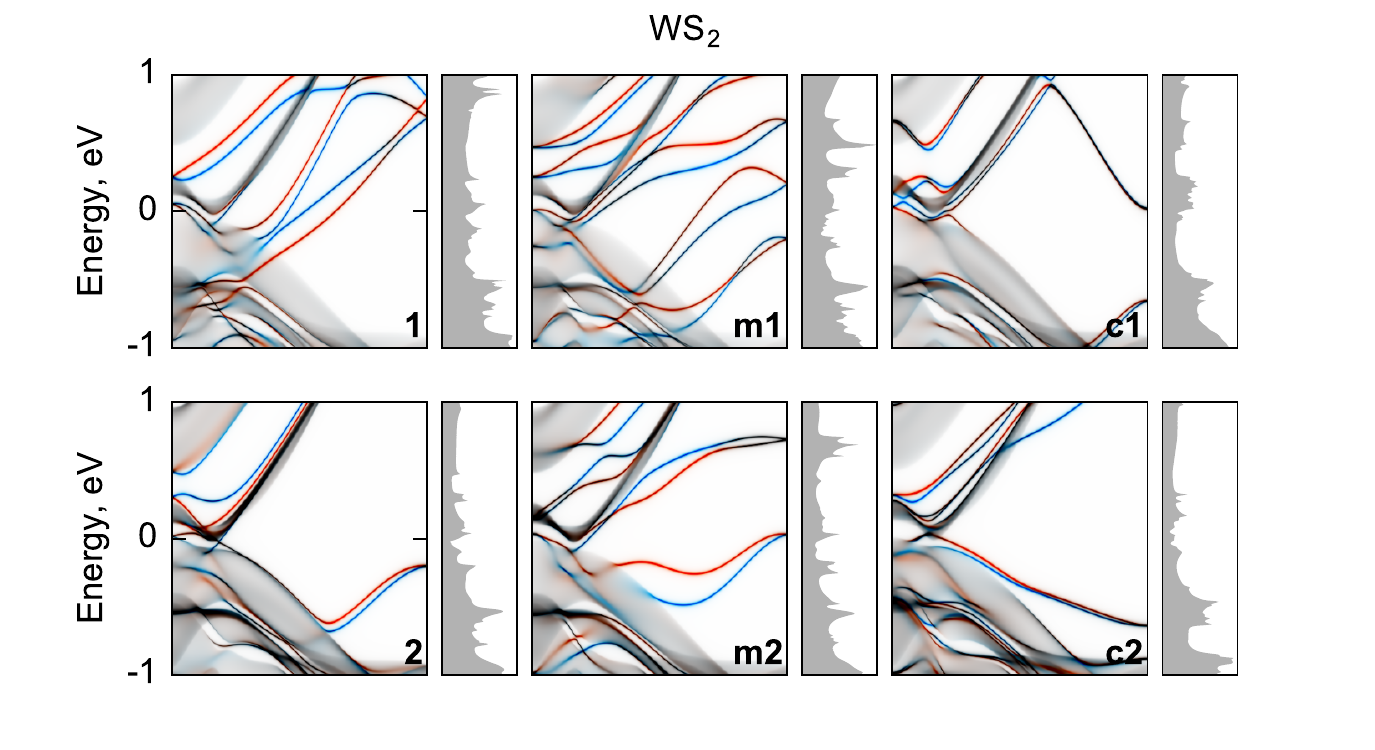}
\caption{Momentum- and spin-resolved local densities of states of six terminations of monolayer 1T'-MoS$_2$ and 1T'-WS$_2$.
$E = 0$ corresponds to the Fermi energy and half of the one-dimensional Brillouin zone is displayed.
The gray areas correspond to the projection of spin-degenerate bulk states, while the edge states are depicted in red and blue for spin-up and spin-down channels, respectively.
Momentum-integrated local density of states are shown on the right. The peaks in these plots correspond to the extrema of edge-state bands providing signatures for the spectroscopic identification of edge terminations. 
}
\label{fig:s2}
\end{figure*}

\begin{figure*}
\includegraphics[width=17cm]{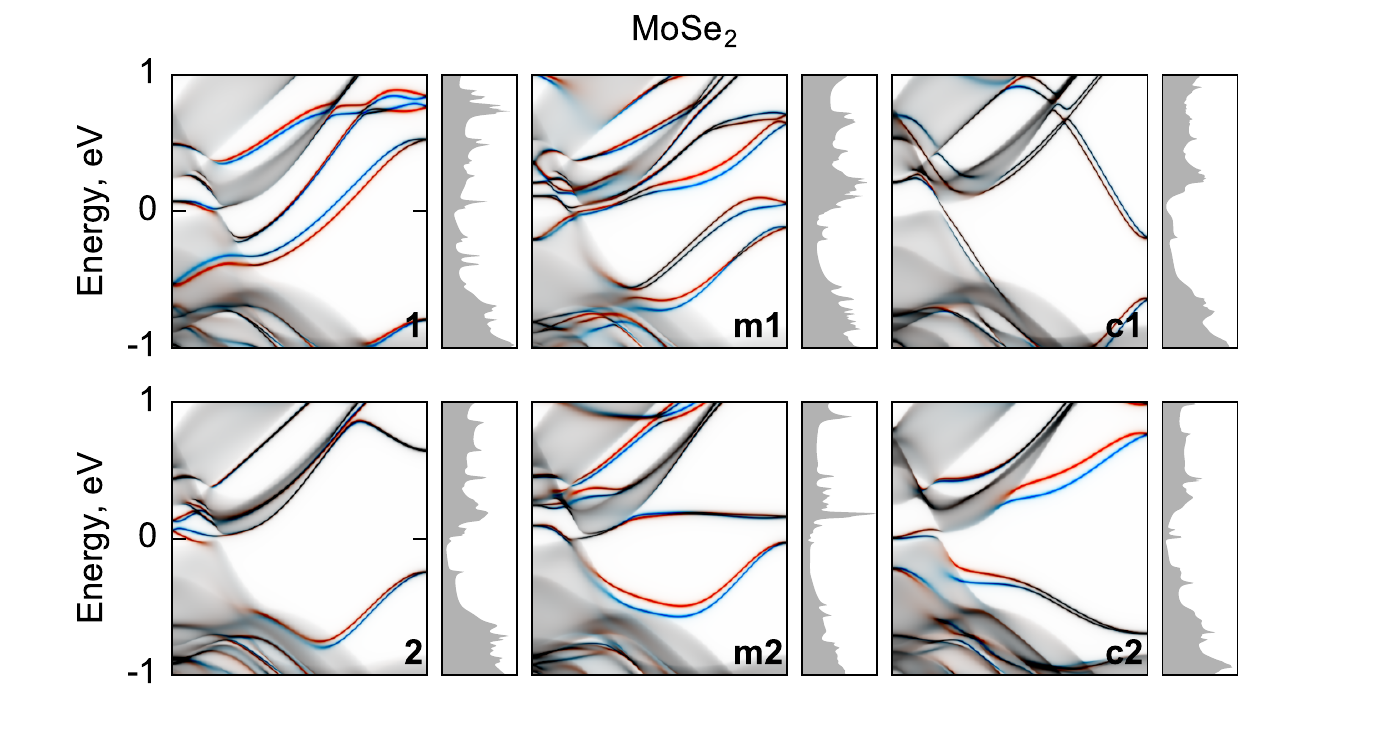}
\includegraphics[width=17cm]{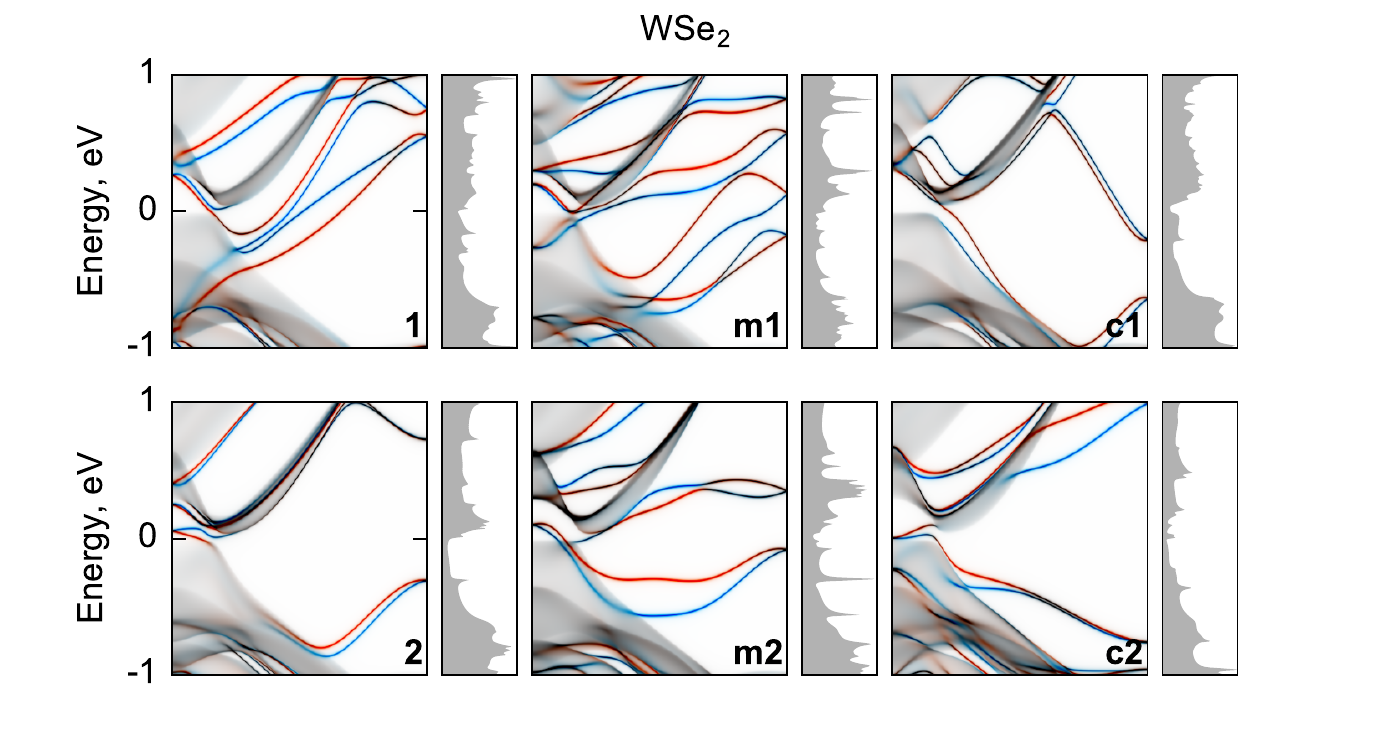}
\caption{Momentum- and spin-resolved local densities of states of six terminations of monolayer 1T'-MoSe$_2$ and 1T'-WSe$_2$.
$E = 0$ corresponds to the Fermi energy and half of the one-dimensional Brillouin zone is displayed.
The gray areas correspond to the projection of spin-degenerate bulk states, while the edge states are depicted in red and blue for spin-up and spin-down channels, respectively.
Momentum-integrated local density of states are shown on the right. The peaks in these plots correspond to the extrema of edge-state bands providing signatures for the spectroscopic identification of edge terminations. 
}
\label{fig:s3}
\end{figure*}

\begin{figure*}
\includegraphics[width=17cm]{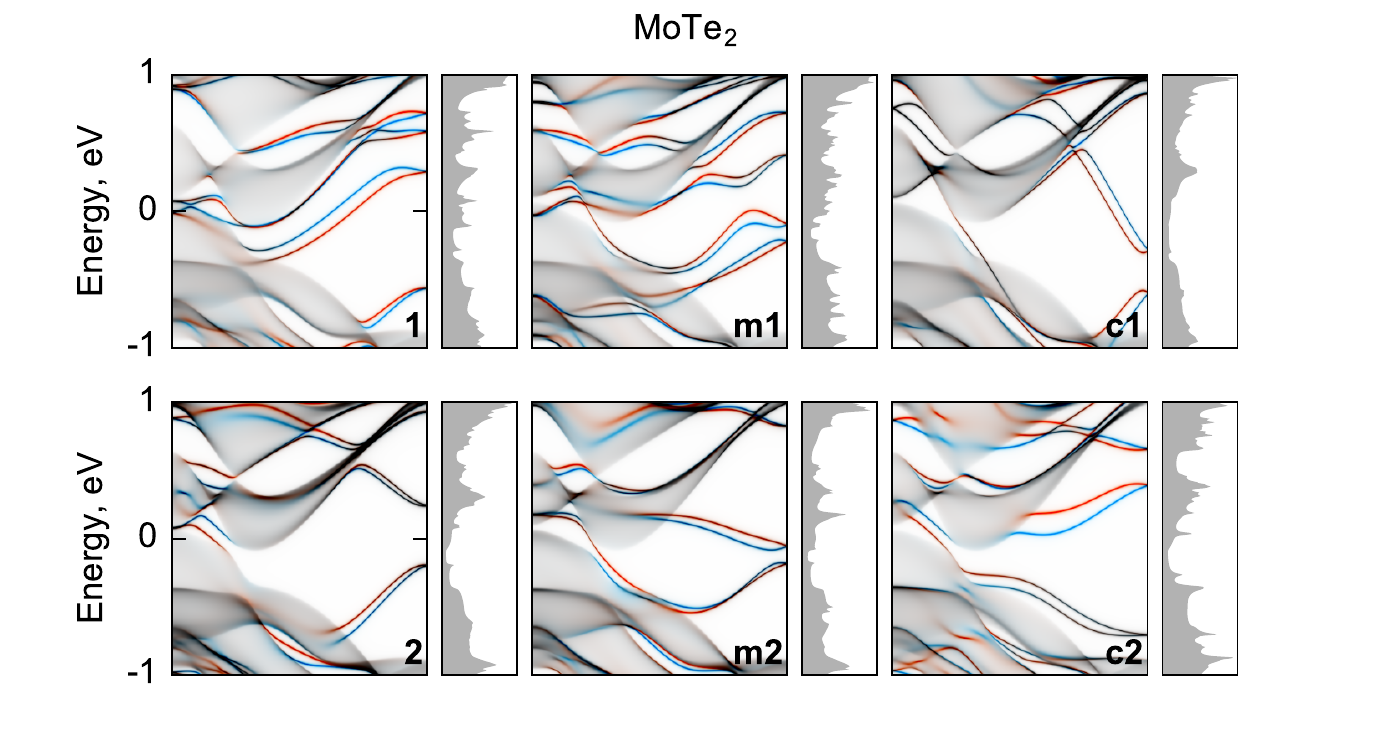}
\includegraphics[width=17cm]{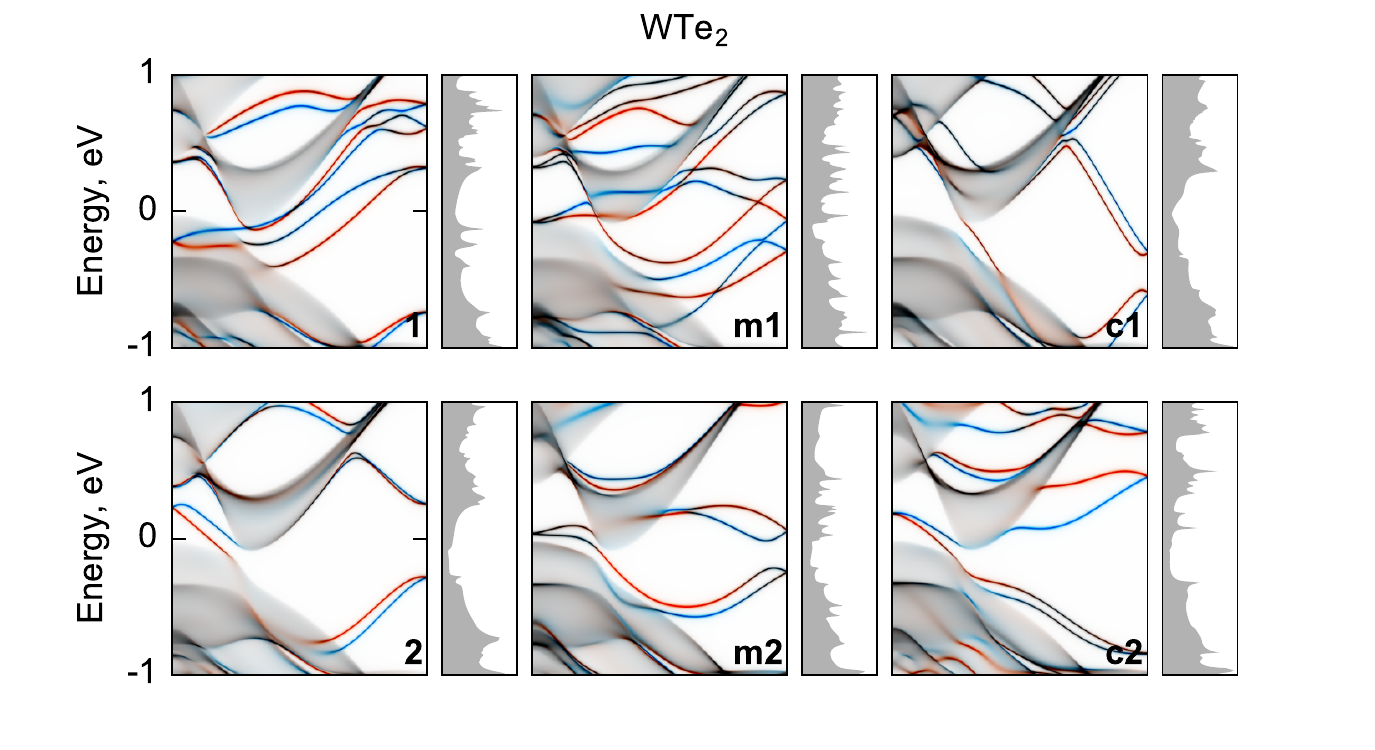}
\caption{Momentum- and spin-resolved local densities of states of six terminations of monolayer 1T'-MoTe$_2$ and 1T'-WTe$_2$.
$E = 0$ corresponds to the Fermi energy and half of the one-dimensional Brillouin zone is displayed.
The gray areas correspond to the projection of spin-degenerate bulk states, while the edge states are depicted in red and blue for spin-up and spin-down channels, respectively.
Momentum-integrated local density of states are shown on the right. The peaks in these plots correspond to the extrema of edge-state bands providing signatures for the spectroscopic identification of edge terminations. 
}
\label{fig:s3}
\end{figure*}
